\documentclass{appolb}
\usepackage{epsfig}
\def\dilog{{\mbox {Li}}_2\,}
\def\xb{{\bar{x}}}
\def\xbar{{\bar{x}}}

\def\eps{\epsilon}
\def\lnlamg{{\log\lambda_G}}
\begin{document}
\date{}
\author{M. Je\.zabek$^{a,b}$ and P. Urban$^{a}$}
\title{One-loop QCD corrected distribution in $B\rightarrow Xe\bar{\nu _e}$ \footnote{Work supported in part by KBN grants 2P03B08414 and 2P03B14715.}.}
\maketitle
\par
\makebox{$^a$} Department of Field Theory and Particle Physics, University of Silesia, Uniwersytecka 4, PL-40007 Katowice, Poland,
\par
\makebox{$^b$} Institute of Nuclear Physics, Kawiory 26a, PL-30055 Cracow, Poland. 
\vspace{1cm}
\par
 We give an analytic formula for the double distribution of hadronic
 invariant mass and charged lepton energy to one-loop order of the
 perturbative QCD. Although infrared singular, this quantity is closely related to physical observables that can be obtained thereof through proper convolution.  
\section{Introduction}
Today, the interest in bottom meson decays has never been stronger.
 Of those, the semileptonic kind are particularly fit for precision measurements of 
the Standard Model parameters such as the Cabibbo-Kobayashi-Maskawa matrix elements $|V_{cb}|$ and $|V_{ub}|$ \cite{LHCB,BaBar} or the quark masses. The theoretical predictions
rely on perturbation theory as well as the effective theory of heavy quarks 
\cite{IW,EH,Georgi,Grinstein} for inclusion of QCD effects.
 The present techniques are able to provide successful description of the semileptonic decays and it has been known that significant contributions to the $B$ decays 
 come from both perturbative \cite{JK,JM,CJK,CJK2,FLS} and nonperturbative \cite{BKPS,Koyrakh,JU} corrections.
\par
In this paper, we give the one-loop perturbative correction to the double differential distribution in terms of the hadron system mass and the charged lepton energy. The masses of the final particles are set to zero throughout the present calculation.
\par
The double differential distribution contains an explicit dependence on the 
fictitious gluon mass as it must due to the infrared singularities present 
both in the virtual and the real radiation corrections. It is not until after 
one integrates over the invariant mass of hadrons that the apparent divergence
cancels to leave a finite, measurable result.
 The distribution we give can be integrated over any desired range of the hadronic mass to yield an experimentally valid quantity. Hence, arbitrary cuts on the hadronic mass may easily be imposed.  
\par
 The result we obtain is consistent with the published formulae for the single
differential distribution in both variables considered. In particular, we have
checked they agree with the charged lepton energy distribution, first published by
Je\.zabek and K\"uhn\cite{JK}, and the hadronic mass distribution, as
found in \cite{FLS,Neubert}.
\section{Kinematics}
\subsection{Variables}
In our treatment of the decay of a $B$ meson, we need to consider both three-
and four-body final states. The latter includes a gluon radiated by the 
initial or final quark. Denote the four-momenta of the involved particles 
as $Q,q,l,\nu,G$ for the $b$ quark, final state quark, electron, neutrino and gluon,
respectively. Also define the hadronic four-momentum $P=q+G$. 
We are working in the rest frame of the $B$ meson, which in the
present approximation reduces to that of the $b$ quark.
\par
The variables we use are $x$ for the electron energy, $t$ denoting the
 invariant mass of the leptons, and $z$ for the  
invariant mass of the hadron system, all scaled according to the following formulae:
\begin{equation}
x=\frac{2(Q\cdot l)}{m_b^2},\qquad t=\frac{(l+\nu)^2}{m_b^2},\qquad z=\frac{(q+G)^2}{m_b^2},
\end{equation}
with $m_b$ the bottom quark mass. Also we will use the scaled masses of the gluon and the final $u$ quark, which both pop up in the intermediate formulae in 
the course of calculating the contributions from three- and four-body processes.
\begin{equation}
\lambda _G=\frac{m_G}{m_b},\qquad \eps = \sqrt{\rho}=\frac{m_u}{m_b}.
\end{equation}
The system of the  quark in the final state and the real gluon is described by the following 
quantities:
\begin{eqnarray}
P_{0}&=&{1 \over 2}(1-t+z)\\
P_{3}&=&\sqrt{P_{0}^{2}-z}={1 \over 2}[1+t^{2}+z^{2}-2(t+z+tz)]^{1 \over 2},\\
P_{\pm}(z)&=&P_{0}(z)\pm P_3(z),\\
{\cal Y}_p&=&{1 \over 2}\ln{{P_+(z)} \over {P_-(z)}}=\ln{{P_+(z)} \over {\sqrt{z}}}
\end{eqnarray}
where $P_0(z)$ and $P_3(z)$ are the energy and the length of the momentum vector
of the system in the $b$ quark rest frame, and ${\cal Y}_p(z)$ is the corresponding 
rapidity. Similarly for the virtual boson $W$:
\begin{eqnarray}
W_0(z)={1 \over 2}(1+t-z),\\
W_3(z)=\sqrt{W_0^2-t}={1 \over 2}[1+t^2+z^2-2(t+z+tz)]^{1/2},\\
W_{\pm}(z)=W_0(z)\pm W_3(z),\\
{\cal Y}_w(z)={1 \over 2}\ln{{W_+(z)} \over {W_-(z)}}=\ln{{W_+(z)} \over \sqrt{t}}.
\end{eqnarray}
 Kinematically, the three body decay is a special case of the four body one, 
with the four-momentum of the gluon set to zero, thus resulting in simply
replacing $z=\rho$. The following variables are then useful:
\begin {eqnarray}
p_0=P_0(\rho)={1 \over 2}(1-t+\rho)&,& p_3=P_3(\rho) = 
\sqrt{p_0^2-\rho},\\
p_{\pm}=P_{\pm}(\rho) = p_0\pm p_3&,&  w_{\pm}=W_{\pm}(\rho) = 1-p_{\mp},\\
Y_p={\cal Y}_p(\rho) = {1 \over 2}\ln{{p_+} \over {p_-}}&,&  Y_w={\cal 
Y}_w(\rho) = {1 \over 2}\ln{{w_+} \over {w_-}}. 
\end{eqnarray}
\subsection{Kinematical boundaries}
The electron energy may vary within the range of
\begin{equation}
0\le x\le 1,
\end{equation}
while for a fixed value of $x$, the allowed interval of $z$ is
\begin{equation}
0 \le z \le 1-x.
\end{equation}
On the other hand, without fixing $x$, the hadronic mass may vary as 
\begin{equation}
0 \le z \le 1,
\end{equation}
and then the following boundaries on $x$ are found for a specified $z$:
\begin{equation}
0 \le x \le 1-z.
\end{equation}
Given values of $x$ and $z$, the intermediate boson invariant mass $t$ 
varies within the following limits,
\begin{equation}
0 \le t \le x(1-\frac{z}{1-x})\equiv t_m.
\end{equation}
\section{Evaluation of the QCD corrections}
The differential decay rate for the process $b\rightarrow u\,l\,\bar{\nu _l}$ can be written as
\begin{equation}
d\Gamma=d\Gamma _0 +d\Gamma _{1,3}+d\Gamma _{1,4}\,,
\end{equation}
where
\begin{equation}\label{gamma0}
d\Gamma _0 = G_F^2 m_b^5|V_{ub}|^2{\cal M}_{1,3}d{\cal R}_3(Q;q,l,\nu)/\pi ^5
\end{equation}
stands for the Born approximation,
\begin{equation}\label{gamma13}
d{\Gamma}_{1,3}={2 \over 3}{\alpha}_sG_F^2m_b^5|V_{ub}|^2{\cal 
M}_{1,3}d{\cal R}_3(Q;q,l,\nu)/{\pi}^6 \end{equation}
comes from the interference between the virtual gluon and Born amplitudes,
and
\begin{equation}
d{\Gamma}_{1,4}={2\over 3}{\alpha}_sG_F^2m_b^5|V_{ub}|^2{\cal 
M}_{1,4}d{\cal R}_4(Q;q,G,l,\nu)/{\pi}^7 \end{equation}
is due to the real gluon emission.  The Lorentz 
invariant $n$-body phase space is defined as 
\begin{equation}
d{\cal R}_n(P;p_1,...,p_n)={\delta}^{(4)}(P-\sum{p_i}){\prod}_i{{d^3{\bf p}_i} \over {2E_i}}.
\end{equation}  
\subsection{Three-body contributions}
The Born approximation is evaluated according to (\ref{gamma0}) with the matrix
element
\begin{equation}
{\cal M}_{0,3}=q\cdot l\,Q\cdot\nu\,,
\end{equation}
and the Dalitz parametrization of the phase space,
\begin{equation}
d{\cal R}_3(Q;q,l,\nu)=\frac{\pi^2}{4}dx\,dt.
\end{equation}
Subsequently, the dot products appearing in ${\cal M}_{0,3}$ are put in terms 
of the variables $x$ and $t$,
\begin{equation}
\begin{array}{llllll}
Q\cdot q&=&{m_b^2\over2}(1+\rho-t),\qquad &l\cdot q&=&{m_b^2\over2}(x-t) \\
Q\cdot \nu &=&{m_b^2\over 2}(1-\rho-x+t),\qquad &l\cdot \nu &=& {m_b^2\over2}t \\
Q\cdot l&=&{m_b^2\over2}x,&\nu\cdot q&=&{m_b^2\over2}(1-\rho-x).
\end{array}
\end{equation}
which readily yields the double differential distribution. On integration over
$t$ the latter yields the Born approximation to the electron energy spectrum (see Eqs. (\ref{LoopResult},\ref{BornTerm})).
\par
The virtual correction is calculated along the same lines, now with the matrix
element 
\begin{eqnarray}\label{M13}
{\cal M}_{1,3}=-\Big[ H_0 q\cdot{l} Q\cdot\nu +H_+\rho Q\cdot\nu Q\cdot{l} +H_- q\cdot\nu q\cdot{l}  \nonumber  \\
+{1 \over 2}{\rho}(H_++H_-)\nu\cdot{l} +{1\over 
2}\rho(H_+-H_-+H_L)[{l}\cdot{(Q-q-\nu)}] (Q\cdot\nu)\nonumber\\ 
-{1\over 2}H_L[{l}\cdot{(Q-q-\nu)}] (q\cdot\nu) \Big],
\end{eqnarray}
substituted into (\ref{gamma13}).
In (\ref{M13}), the form factors are as follows,
\begin{eqnarray}
\nonumber H_0=4(1-Y_pp_0/p_3)\ln{\lambda}_G+(2p_0/p_3)[\dilog(1-{{p_-w_-}\over {p_+w_+}})\\
\nonumber -\dilog(1-{{w_-}\over {w_+}}-Y_p(Y_p+1)+2(\ln\sqrt{\rho}+Y_p)(Y_w+Y_p)]\\
+[2p_3Y_p+(1-\rho-2t)\ln\sqrt{\rho}]/t+4,\\
H_{\pm}={1\over 2}[1\pm (1-\rho)/t]Y_p/p_3\pm {1\over t}\ln\sqrt{\rho},\\
H_L={1\over t}(1-\ln\sqrt{\rho})+{{1-\rho}\over {t^2}}\ln\sqrt{\rho}+{2\over {t^2}}Y_pp_3+{{\rho} \over t}{{Y_p} \over {p_3}}.
\end{eqnarray}
Integration over the invariant mass of the intermediate $W$ boson gives the 
desired contribution to the double differential distribution,
\begin{eqnarray}
\frac{d\Gamma_{1,3}}{12\Gamma_0\,dxdz}\!\!\!\!&=&\!\!\!\!\delta(z)\big\{  {1\over3}\lnlamg \big[ 10x 
- 25x^2 + {34\over3}x^3 
          +( 10 - 24x + 18x^2 - 4x^3)\log\xb \nonumber\\
&&        - 6x^2\log\eps + 4x^3\log\eps \big]
        +{1\over3}(2x^3-3x^2)\dilog(x)
        +{1\over18}\log\xb\big[121\nonumber\\
&&-276x+195x^2-40x^3           -\log\xb(30-72x+54x^2-12x^3 )\big] \nonumber\\
&&  + {121\over18}x- {443\over36}x^2+ {128\over27}x^3 +{1\over6}\log\eps(-3x^2+2x^3+6x^2\log\eps\nonumber\\
&&-4x^3\log\eps)\big\},
\end{eqnarray}
where we have denoted
\begin{equation}
\xb=1-x.
\end{equation}
While we present the decay rate assuming massless final particles, it is not
possible to eliminate the final quark as well as gluon mass dependence out of
 the matrix element alone. This is removed once the integrated real radiation term is added. 

\subsection{Four-body contributions}
The four-body phase-space is decomposed as follows:
\begin{equation}
d{\cal R}_4(Q;q,G,l,\nu)=dz\,d{\cal R}_3(Q;P,l,\nu)\,d{\cal R}_2(P;q,G).
\end{equation}
Discarding terms vanishing along with the gluon mass $\lambda _G\rightarrow 0$, the matrix element representing the real gluon radiation may be  written as
\begin{equation}
{\cal M}_{1,4}={{{\cal B}_1}\over {(Q\cdot G)^2}}-{ {{\cal B}_2}\over {Q\cdot G P\cdot G}}+{ {{\cal B}_3} \over {(P\cdot G)^2}},
\end{equation}
where
\begin{eqnarray}
{\cal B}_1=q\cdot {l}[Q\cdot\nu(Q\cdot G-1)+G\cdot\nu-Q\cdot\nu Q\cdot G],\\
{\cal B}_2=q\cdot {l}[G\cdot\nu -q\cdot\nu Q\cdot G+Q\cdot\nu(q\cdot G-Q\cdot G-2q\cdot Q)]\nonumber\\
+Q\cdot\nu(Q\cdot{l} q\cdot G-G\cdot{l} q\cdot Q),\\
{\cal B}_3=Q\cdot\nu(G\cdot{l} q\cdot G-\rho{l}\cdot P).
\end{eqnarray}
We refer the reader to \cite{CJK} for the details of the integration
that leads to the triple differential rate in terms of $x,t$ and $z$. This rate
is conveniently split into terms according as they are infrared convergent or
 divergent. Hence we can write,
\begin{equation}
\frac{d\Gamma_{1,4}}{12\Gamma _0\,dx\,dt\,dz}={\cal F}_{conv}+{\cal F}_{div}.
\end{equation}
This expression is subsequently integrated over $t$.
The infrared finite part has been integrated with the help of FORM. 
The formulae obtained in this way suffer from infrared and collinear
divergences. Thence the gluon and final quark masses  
subsist as regulators in spite of the limit we have taken. Of course,
both remnant dependences vanish after the integration over the hadronic
system mass is performed, which involves summing the virtual and real
contributions.

\section{Results}
The first order QCD corrected double differential decay rate can be written in
the form,
\begin{equation}\label{LoopResult}
\frac{d\Gamma}{12\Gamma _0\,dx\,dz}=f_0(x)\delta (z)+\frac{2\alpha _s}{3\pi}\big[f_1^s \delta(z)+f_1^c(x,z) \theta(z-\lambda _G^2)\big],
\end{equation}
with
\begin{equation}
\Gamma _0=\frac{G_F^2 m_b^5}{192 \pi^3},
\end{equation}
where the first term on the right hand side is the Born approximation, given by
\begin{equation}\label{BornTerm}
f_0(x)={1\over6}x^2(3-2x),
\end{equation}
while
\begin{eqnarray}\label{f1s}
f_1^s(x)&=&-4f_0(x)\log^2\lambda_G+\frac{1}{18}(120x-291x^2+130x^3)\lnlamg\nonumber\\
&& +{1\over3}(-10+24x-18x^2+4x^3)\log\xb \log\left(\frac{\xbar}{\lambda_G^2}\right)\nonumber\\
&&      +{1\over6}(83-196x+145x^2-32x^3)\log\xb\nonumber\\
&&      +{1\over{18}}(249x-426x^2+155x^3)
        -{2\over3}f_0(x)\left[\pi^2+3\dilog(x)\right],
\end{eqnarray}
and
\begin{eqnarray}\label{f1c}
f_1^c(x,z)&=&{1\over z}\big[-2f_0(x)\log z+{1\over36}\left(120x-291x^2+130
x^3\right)+{1\over3}(10-24x\nonumber\\
&&+18x^2-4x^3)\log\xb\big]+\big[(10z+3z^2)/\xb-(z+2z^2)/\xb^2\nonumber\\
&&+{2\over3}z^2/\xb^3\big]\log\frac{z}{\xb^2}
       + {1\over2}\xb^{-1}  (  - 69z - 25z^2 )
       + {1\over4}\xb^{-2}  (   11z + 25z^2 )\nonumber\\
&& 
       + \xb^{-3}  ( - 29z^2/18 )+  ( 101/2  - 17xz - 81x  + 45x^2/2\nonumber\\
 && - 31z - 3z^2/2 ) \log\xb
       +  (   32 - 6xz - 30x + 4x^2 + 29z\nonumber\\ 
&&      + z^2 )\log\xb \log\frac{z}{\xb}+\log z (21xz+32x-16x^2-9z-5z^2/3)\nonumber\\
&&       
         - 20xz + 101x/2  - 97x^2/4  + 127z/4 + 283z^2/36.
\end{eqnarray}
The above formula is easily integrated over either of the variables
 to give the single differential distributions in hadronic system 
mass or charged lepton energy.
Then expressions confirming previous calculations
\cite{JK,FLS,Neubert}  are found. 
The evident singularity of this distribution disappears after
 integration over the hadronic system mass. That this indeed is so,
 can be seen by expressing it in terms of the following distributions,
\begin{eqnarray}
\left( \frac{1}{z}\right)_+&=&\lim_{\lambda\rightarrow 0} \left
(  \frac{1}{z}\theta(z-\lambda) + \log \lambda \,\delta(z)\right),\\
\left( \frac{\log z}{z} \right)_+ &=& \lim_{\lambda\rightarrow 0}
\left(\frac{\log z}{z} \theta(z-\lambda) +\frac{1}{2}\log^2\!\lambda\, \delta(z)\right).
\end{eqnarray}
The substitution of these to Eqs.(\ref{f1s},\ref{f1c}) results in the formal
identification,
\begin{eqnarray}\label{Drules}
\theta(z-\lambda_G^2)\frac{1}{z}&=&\theta(z-\lambda_G^2)\left
( \frac{1}{z}\right)_+ - \delta(z)\log\lambda_G^2,\\
\label{DruleB}\theta(z-\lambda_G^2)\frac{\log z}{z}&=&\theta(z-\lambda_G^2)\left
( \frac{\log z}{z} \right)_+ - {1\over2}\log^2\lambda_G^2\delta(z).
\end{eqnarray}
Upon application of Eqs. (\ref{Drules},\ref{DruleB}) to the correction terms, the
latter take on the following form:
\begin{eqnarray}\label{f1sD}
f_1^s(x)&=& {1\over3}(-10+24x-18x^2+4x^3)\log^2\xb \nonumber\\
&&      +{1\over6}(83-196x+145x^2-32x^3)\log\xb\nonumber\\
&&      +{1\over{18}}(249x-426x^2+155x^3)
        -{2\over3}f_0(x)\left[\pi^2+3\dilog(x)\right],
\end{eqnarray}
\begin{eqnarray}\label{f1cD}
f_1^c(x,z)&=&-2f_0(x)\left( \frac{\log z}{z} \right)_+ +\left
( \frac{1}{z}\right)_+\big[{1\over36}\left(120x-291x^2+130 
x^3\right)\nonumber\\
&&+(\frac{10}{3}-8x+6x^2-\frac{4}{3}x^3)\log\xb\big]+\big[(10z+3z^2)/\xb-(z+2z^2)/\xb^2\nonumber\\
&&+{2\over3}z^2/\xb^3\big]\log\frac{z}{\xb^2}
       + {1\over2}\xb^{-1}  (  - 69z - 25z^2 )
       + {1\over4}\xb^{-2}  (   11z + 25z^2 )\nonumber\\
&& 
       + \xb^{-3}  ( - 29z^2/18 )+  ( 101/2  - 17xz - 81x  + 45x^2/2\nonumber\\
 && - 31z - 3z^2/2 ) \log\xb
       +  (   32 - 6xz - 30x + 4x^2 + 29z\nonumber\\ 
&&      + z^2 )\log\xb \log\frac{z}{\xb}+\log z (21xz+32x-16x^2-9z-5z^2/3)\nonumber\\
&&       
         - 20xz + 101x/2  - 97x^2/4  + 127z/4 + 283z^2/36.
\end{eqnarray}

\begin{figure}[t]
\epsfig{file=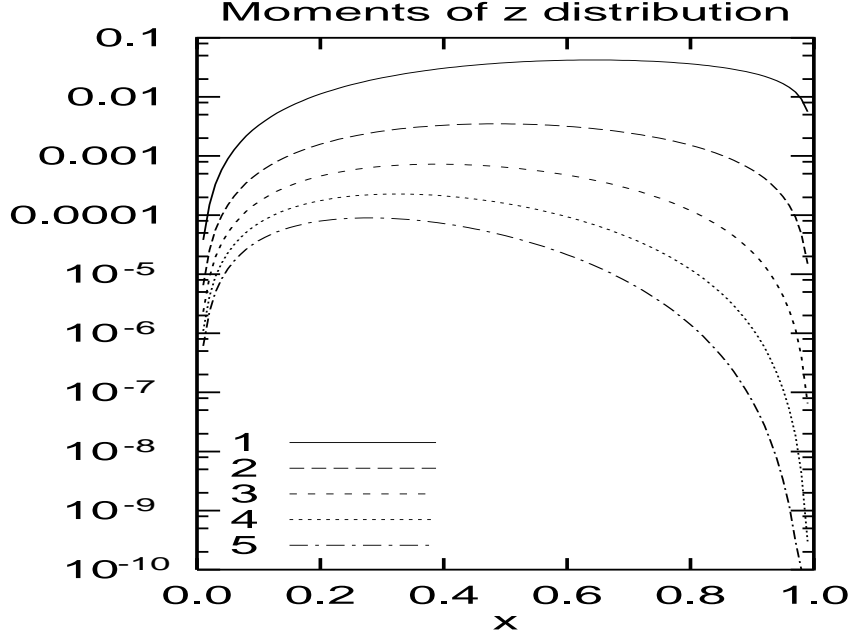,height=250pt,width=350pt}
\caption{\label{FigMoments}Moments of hadron mass distribution $m_n$ as a function of the scaled
electron energy $x$. The first five moments are shown as indicated by the legend.}
\end{figure}

Clearly, the gluon mass does not enter the integrated distribution, defined as
\begin{equation}
F(x,z)={1\over{12\Gamma
_0}}\int_0^{z}dz'{\frac{d\Gamma}{dx\,dz'}}=f_0(x)
+\frac{2\alpha_s}{3\pi}F_1(x,z)\,,
\end{equation}
for which we obtain,
\begin{equation}
F_1(x,z)=(c_1+ c_2\,z +c_3\,z^2+c_4\,z^3)\log z 
         +c_5\log ^2 z + c_6 z +c_7 z^2 + c_8 z^3 +c_9.
\end{equation}
The coefficients $c_1$ to $c_9$ are as follows,
\begin{eqnarray}
c_1 &=& {1\over{36}}\big[ ( 120 - 288x+ 216x^2 - 48x^3 )\log\xb 
               + 120x  - 291x^2 + 130x^3 \big],\nonumber\\
c_2 &=& (-30x+ 4x^2+ 32) \log\xb+ 32x- 16x^2,\nonumber\\
c_3 &=& 5{\xb}^{-1}-{1\over2}{\xb}^{-2}+{1\over2}(-6x+29)\log\xb+(21x-9)/2,\nonumber\\
c_4 &=& \xb^{-1}-{2\over3}{\xb}^{-2}+{2\over9}{\xb}^{-3}+{1\over3}\log\xb - 5/9,\nonumber\\
c_5 &=& - x^2/2 + x^3/3,\nonumber\\
c_6 &=& {1\over2}(-102x + 37x^2+ 37)\log\xb +  ( 30x -
        4x^2- 32) \log^2\xb+\nonumber\\
&& (74x- 33x^2)/4,\nonumber\\
c_7 &=& {1\over4}(-28x-91)\log\xb + {1\over2}(6x-29)\log^2\xb
        -10{\xb}^{-1}\log\xb
        -{79\over4}{\xb}^{-1}+\nonumber\\
&& {\xb}^{-2}(\log\xb+{13\over8})- {61\over4}x+
       {145\over8},\nonumber\\
c_8 &=& {1\over{36}}\big\{-22\log\xb-12\log^2\xb-72{\xb}^{-1}\log\xb
        -162{\xb}^{-1}\nonumber\\
&&+48{\xb}^{-2}\log\xb+83{\xb}^{-2}
         -16{\xb}^{-3}\log\xb-22{\xb}^{-3}+101\big\},\nonumber\\
c_9 &=& {1\over{36}}\big\{6  ( 83  - 196x + 145x^2 - 32x^3    )\log\xb
       + 12  (  - 10  + 24x \nonumber\\
&&- 18x^2 + 4x^3   )\log^2\xb       -  12\pi^2 x^2 + 8\pi^2x^3   + 498x
\nonumber\\
&&
        - 852x^2 + 310x^3   - 36x^2\dilog(x) + 24x^3\dilog(x)\big\}.
\end{eqnarray}
One way of making comparison between the parton model predictions and the 
resonance ridden experimental data is to consider moments of distribution.
We define those as
\begin{equation}
M_n(x)={1\over{12\Gamma_0}}\int_0^{1-x}z^n\frac{d\Gamma}{dx\,dz}dz.
\end{equation}
While the zeroth moment corresponds to the electron energy distribution itself,
the singular part of the double distribution leaves no trace on the higher 
moments. In fact, they are then vanishing in the Born approximation. The first
five moments are expressed in terms of the following functions,
\begin{equation}
M_n(x)=\frac{2\alpha_s}{3\pi}m_n(x),\qquad n\ge 1,
\end{equation}
which are given by the formulae beneath (see also Fig. \ref{FigMoments}):
\begin{eqnarray}
m_1 &=&
         (  - 35/144 + 5x^2/8 - 4x^3/9 + x^4/16 ) \log\xbar \nonumber \\
&&       - 35x/144 + 19x^2/72 - x^3/48, \\
m_2 &=&   (  - 449/3600 + 7x/24 - 61x^2/360 - 2x^3/45 + 13x^4/240\nonumber \\
&&
         - 13x^5/1800 )\log\xbar    - 449x/3600 + 533x^2/1800 - 9x^3/40\nonumber \\
&& + 109x^4/1800 - x^5/144,\\
m_3 &=&   (  - 103/1800 + 119x/600 - 29x^2/120 + 19x^3/180 + x^4/
         120\nonumber\\
&& - 3x^5/200 + x^6/600 )\log\xbar      - 103x/1800 + 697x^2/3600\nonumber\\
&& - 653x^3/2700 + 49x^4/360 - 7x^5/200 + 
         47x^6/10800,\\
m_4&=&  \log\xbar  (  - 1313/44100 + 2x/15 - 81x^2/350 + 59x^3/315 - 5x^4/
         84\nonumber\\
&& - x^5/175 + 2x^6/315 - 2x^7/3675 )      - 1313x/44100 + 5729x^2/44100\nonumber\\
&&  - 11x^3/49 + 857x^4/4410  - 781x^5/8820
         + 311x^6/14700\nonumber\\
&& - 19x^7/7350,\\
m_5&=&  \log\xbar  (  - 485/28224 + 23x/245 - 211x^2/1008 + 151x^3/630\nonumber\\
&& 
          - 95x^4/672 + 2x^5/63 + 29x^6/5040 - x^7/294 + 31x^8/141120 )\nonumber\\
&&  - 485x/28224 + 1321x^2/14400 - 15879x^3/78400\nonumber\\
&& + 33599x^4/141120  - 
         4523x^5/28224 + 979x^6/15680\nonumber\\
&& - 9809x^7/705600 + 163x^8/100800.
\end{eqnarray}
\section{Summary}
One-loop QCD correction has been found to the electron energy and
hadronic system mass double distribution in semileptonic $B$
decays. The result has not been published before. It agrees with the
known single distributions in both variables. Although infrared singular,
 it is of use for experimental analysis with
appropriate cuts imposed. 
\pagebreak

\end{document}